\documentclass[12pt]{iopart}

\usepackage{psfig}

\begin{document}

\newcommand{\suin}{\sum \hspace*{-5mm} \int\limits}

\letter{On the role of shake-off in single-photon double ionization}
\author{T Pattard, T Schneider and JM Rost}
\address{Max Planck Institute for the Physics of Complex Systems,
N\"othnitzer Stra{\ss}e 38, D-01187 Dresden, Germany}
\ead{tpattard@mpipks-dresden.mpg.de}

\begin{abstract}
The role of shake-off for double ionization of atoms by a single
photon with finite energy has become the subject of debate. In this letter,
we attempt to clarify the meaning of shake-off at low photon energies by
comparing different formulations appearing in the literature and by suggesting
a working definition. Moreover, we elaborate on the foundation and
justification of a mixed quantum-classical ansatz for the calculation of
single-photon double ionization.
\end{abstract}


\maketitle

\nosections
It is well known that the double-to-single cross section ratio for ionization
of atoms by a single photon does not vanish at high photon energies. Rather, it
approaches a finite constant which can be explained in the framework of a
sudden approximation by the so-called shake-off mechanism. While shake-off
is well defined in the asymptotic high-energy limit, its meaning at finite
energies is less clear and has been the subject of debate recently.
In fact, a number of definitions can be found in the literature,
e.g., \cite{abe69,tho84,hin93,pat01b,khe01,lin02,sch02}. Some of these
definitions are based on formal diagrammatic perturbation expansion techniques
\cite{hin93,khe01}, others on more general physical arguments
\cite{tho84,pat01b}, others on a simple extension of the sudden approximation
idea to finite energies \cite{abe69,lin02,sch02}.
Most of them (with the exception of \cite{hin93}, see below)
have in common that they approach the well-known asymptotic
expression at high energies. At low and intermediate energies, however,
they may differ markedly (e.g., some show a monotonic dependence on energy
while others have a maximum at some finite energy, some even exceed the
total double-to-single cross section ratio measured experimentally).
Thus, so far no unique definition for
shake-off at finite energies exists, and it is not obvious what
``the best'' definition might be.
On the other hand, in particular in connection with the interpretation of
experimental data, attention has been given to
the question what physical mechanisms dominate double ionization and
at which energies the different mechanisms are important \cite{kna02}.

One may argue that a satisfying definition of shake-off
would be one based on physical
principles in addition to mathematical rigor. Hence, a ``good'' definition
should separate shake-off as much as possible from other ionization
mechanisms. Clearly, such a separation is not strictly possible in the
presence of other available routes to ionization and can only
be approximate, which makes the discussion of a shake-off {\em mechanism} a
somewhat delicate issue. In comparing calculations with experiments, e.g.,
one should always keep in mind that no strict one-to-one correspondence
between a shake-off mechanism as an approximate physical picture and a
separate calculation of shake-off can be expected due to the neglect of
interference between possible decay routes. Nevertheless, since such simple
physical pictures, to the extent of their applicability, can be very valuable 
for our intuitive understanding of physical processes, a definition separating
shake-off from ``non-shake-off'' would seem most rewarding conceptually. One
such definition has recently been given by Schneider et al. \cite{sch02}
(hereafter referred to as SCR), where the single-photon double ionization
process has been
described in terms of two separate contributions, namely ``shake-off'' and
``knock-out''. The method used in SCR was shown to lead to excellent
agreement with experiment and {\em ab-initio} calculations for double
ionization from the ground state \cite{sch02,sch03}, and very recently also
from excited states \cite{emm03}, of helium. Thus, it suggests itself as a
more or less natural ``operational'' definition of shake-off in the
framework of the ``half-collision'' picture of single-photon multiple
ionization \cite{sam90,pat01b}.

The calculation reported in SCR starts from a mixed quantum-classical ansatz
that is based on the separation of the photoabsorption process (which is not
treated explicitly in the calculation) from the subsequent evolution of the
system. It treats this evolution (i.e.\ the redistribution of the energy
between the two electrons) in the spirit of an (e,2e)--like process with the
additional possibility of shake-off. Such a ``half-collision'' picture has
been originally suggested by Samson \cite{sam90} and elaborated by Pattard
and Burgd\"orfer \cite{pat01b}, allowing for shake-off processes which
are not taken into account in Samson's original model. In the SCR ansatz,
the (e,2e)--like (``knock-out'') part of the cross section is calculated
using a classical trajectory Monte Carlo method, to which the shake-off as a
purely quantum mechanical process is added on top. In this spirit, shake-off
is introduced as a more or less {\sl ad-hoc} quantum correction to an
essentially classical treatment.
Here, we start from a fully quantum mechanical expression and see which kind
of approximations lead to an SCR-like ansatz. In this way,
further insight into the validity of the ansatz, concerning both technical
details of the calculation as well as the approximate separation of physical
mechanisms (shake-off and knock-out), can be obtained.

In ref.\ \cite{pat01b}, a Born series for the transition amplitude from the
ground state $\psi_i$ to a final state $\psi_f^{(0)}$ of a two-electron target
following single-photon absorption has been derived. It was shown that, under
the assumption of negligible electron--electron correlation in the ionized
final state, the transition amplitude can be written as
\begin{equation}
\label{e1}
a_{fi} = - 2 \pi i \; \delta \left( E_f - E_i - \omega \right) \langle
\psi_f^{(0)} |
\left( 1 - i \int\limits_0^\infty dt \; {\rm e}^{i H_0 t} \; T_{ee} \;
{\rm e}^{- i H_0 t} \right) V_{pe} | \psi_i \rangle \, .
\end{equation}
In the above equation, $V_{pe}$ is the photon-electron interaction, usually
taken in dipole approximation, $H_0$
is the final-state Hamiltonian $H_0 = H_{at} - V_{ee}$ and
$T_{ee}$ denotes the Coulomb $T$-matrix for electron-electron
scattering.
$\psi_f^{(0)}$ is an eigenfunction of $H_0$, i.e.\ a product of two
one-electron states, due to the assumption of vanishing electron-electron
correlation in the final state (where
at least one electron is ionized),
while $\psi_i$ is the fully correlated
initial (ground) state of the target.
Introduction of a complete set of intermediate states then allows for a
separation (on the {\em amplitude} level!) of the initial photon absorption
from the subsequent propagation
\begin{equation}
\label{e2}
a_{fi} = - 2 \pi i \;\; \delta \left( E_f - E_i - \omega \right)\; \suin_a\;
\langle \psi_f^{(0)} | S_+ | \psi_a \rangle \; \langle
\psi_a | V_{pe} | \psi_i \rangle \, ,
\end{equation}
where the notation
\begin{equation}
\label{e3}
S_+ \equiv 1 - i \int\limits_0^\infty dt \; {\rm e}^{i H_0 t}
\; T_{ee} \; {\rm e}^{- i H_0 t}
\end{equation}
is motivated by its resemblance of a conventional scattering $S$-matrix. Note,
however, that $S_+$ is not strictly an $S$-matrix for electron-electron
scattering since the time integral in equation (\ref{e3}) is restricted to
positive $t$, i.e.\ $S_+$ corresponds to a {\em half}-collision.
Furthermore, let us choose the complete set $\{ \psi_a \}$ in such a way that
\begin{equation}
\label{e4}
\psi_{abs} (1,2) \equiv \frac{\left( V_{pe} \psi_i \right)}{\sqrt{\langle
(V_{pe} \psi_i) | (V_{pe} \psi_i) \rangle}}
\end{equation}
is contained in this set. From the orthogonality condition for the basis
states, all other basis states are then orthogonal to $V_{pe} \psi_i$. Thus
the sum over intermediate states in equation
(\ref{e2}) collapses to a single term
and we can write
\begin{equation}
\label{e5}
a_{fi} = - 2 \pi i \;\; \delta \left( E_f - E_i - \omega \right)\; \langle
\psi_f^{(0)} | S_+ | \psi_{abs} \rangle \; \langle
\psi_{abs} | V_{pe} | \psi_i \rangle \, .
\end{equation}
The photon absorption probability is then given by a sum over all final states
$\psi_f^{(0)}$ of the transition probability per unit time into the state
$\psi_f^{(0)}$
\begin{equation}
\label{e6}
P_{abs} = 2 \pi \sum\limits_f \delta \left( E_f - E_i - \omega \right)\;
| \langle \psi_f^{(0)} | S_+ | \psi_{abs} \rangle |^2 \; | \langle
\psi_{abs} | V_{pe} | \psi_i \rangle |^2 \, .
\end{equation}
On the other hand, it is also directly given by
\begin{equation}
\label{e7}
P_{abs} = 2 \pi \sum\limits_f \delta \left( E_f - E_i - \omega \right)\;
| \langle \psi_f | V_{pe} | \psi_i \rangle |^2 \, .
\end{equation}
(Note that the $\psi_f$ in equation (\ref{e7}) are eigenfunctions of the
full atomic Hamiltonian $H_{at}$ including electron-electron interaction,
in contrast to the final states in equation (\ref{e6}).)
From equation (\ref{e4}), however, it immediately follows that
\begin{equation}
\label{e8}
| \langle \psi_{abs} | V_{pe} | \psi_i \rangle |^2
=
\langle V_{pe} \psi_i | V_{pe} \psi_i \rangle
=
\sum\limits_f | \langle \psi_f | V_{pe} | \psi_i \rangle |^2
\, ,
\end{equation}
which in general does not coincide with the expression (\ref{e7}) involving
an additional delta function. Hence, it can be seen that it is precisely the
off-shell (i.e.\ off the final-state energy shell) part of $\psi_{abs}$ which
prohibits an exact factorization of the transition probability into a
photon absorption probability and an ``energy redistribution'' part. Note that
this can also be seen from equation
(\ref{e2}) directly if the set of intermediate
states $\psi_a$ is chosen as eigenstates of $H_0$. Then
\begin{equation}
\label{e9}
\langle \psi_f^{(0)} | S_+ | \psi_a \rangle = \delta_{fa} -
i \int\limits_0^\infty
dt \; {\rm e}^{i (E_f - E_a) t} \; \langle \psi_f^{(0)} | T_{ee} |
\psi_a \rangle
\end{equation}
and from
\begin{equation}
\label{e10}
\int\limits_0^\infty dt \; {\rm e}^{i (E_f - E_a) t} = \pi \delta \left(
E_f - E_a \right) + i \frac{\cal{P}}{E_f - E_a}
\end{equation}
it becomes clear that the off-shell part of $\psi_{abs}$ is a consequence
of time ordering \cite{mcg03}, i.e.\ that requiring the photon to be
absorbed first restricts the time integral in (\ref{e3}) to positive $t$.

For the remainder of this discussion, let us neglect the photon absorption
process and focus on the second step of the ionization process, namely the
``half-collision'' part of equation (\ref{e5})
\begin{equation}
\label{e11}
a_{f,abs} \equiv \sqrt{2 \pi} \; \delta \left(E_f - E_i - \omega \right)
\langle \psi_f^{(0)} | 1 - i \int\limits_0^\infty dt \; {\rm e}^{i H_0 t} \;
T_{ee} \; {\rm e}^{- i H_0 t} | \psi_{abs} \rangle \; .
\end{equation}
(The splitting of the factor $2 \pi$ is motivated by the
fact that the resulting shake {\em probability} to be discussed below reduces
to the correct asymptotic form at high energies.)
$a_{f, abs}$ is seen to consist of two parts, namely the interaction free
unity operator ``1'' and the operator ``$T$'' involving
electron-electron interaction.
Naturally, the former can be associated with a shake process while the latter
corresponds to a ``knock-on'' (we use the
expressions shake and knock-{\sl on} for any final state and the terms
shake-{\sl off} and knock-{\sl out} for doubly ionized states as in SCR).
Hence, we propose
\begin{equation}
\label{e12}
a_{f,abs}^{S} \equiv \sqrt{2 \pi} \; \delta \left(E_f - E_i - \omega \right)
\langle \psi_f^{(0)} | \psi_{abs} \rangle
\end{equation}
as a working definition for the shake amplitude at a finite excess energy
$E = E_i + \omega$.
However, the shake and knock-on contributions are summed on the {\sl amplitude}
level. To
arrive at the SCR ansatz, the additional approximation of an incoherent
summation of shake and knock-on has to be made\footnote{In addition to that,
the knock-on part has been obtained from a classical CTMC calculation in SCR.
While such a treatment is frequently employed in the study of atomic
collision processes, an evaluation of its quality is beyond the scope of the
present Letter.}. The error
introduced by this approximation is at most of the order of the smaller of the
two contributions, i.e.\ it goes to zero in the high- as well as low-energy
limit and could only contribute significantly at intermediate energies. Even
there it was found in SCR that the error is of the order of a few
percent only (at least for the double-to-single ionization ratio of helium).
One would speculate \cite{pat01b,sch03}
that this is
to a large extent due to the
population of different final states by the two mechanisms. For shake,
e.g., the ``shaken'' electron will be in an s-state, while the knock-on
mechanism will also populate higher angular momentum states. Calculations of
angular-differential cross sections should shed further light on this
question.

From equation (\ref{e12}), the {\em probability} for a shake process to a final
state $\psi_f^{(0)}$ per unit time is found to be
\begin{equation}
\label{e13}
P_{f,abs}^{S} = \delta \left( E_f - E \right) \; | \langle \psi_f^{(0)} |
\psi_{abs}
\rangle |^2 \; .
\end{equation}
With the definition of $\psi_{abs}$, equation (\ref{e4}),
this is more explicitly
written in terms of the initial state as
\begin{equation}
\label{e14}
P_{f,abs}^{S} = \delta \left( E_f - E \right) \; \frac{| \langle \psi_f^{(0)} |
V_{pe} \psi_i \rangle |^2}{\langle V_{pe} \psi_i | V_{pe} \psi_i \rangle} \; .
\end{equation}
This expression differs somewhat from the one given by {\AA}berg
\cite{abe69} (equation (7) of SCR). In contrast to the former, it contains the
``photoabsorption operator'' $V_{pe}$. It seems that the current definition
equation (\ref{e14}) is preferable since it arises naturally from the
preceding arguments: The sudden approximation underlying the shake-off
picture is with respect to the electron-electron interaction, not with
respect to the photon absorption.  As noted by SCR,
if the photoelectron is in an s-state initially, it will be in a p-state after
absorption of the photon. That is, to the extent that the dipole approximation
is valid for the photon-electron interaction and that the single-particle
angular momentum $l$ is a good quantum number, the $\psi^\nu$ defined by
{\AA}berg is identically zero. It should be noted, however, that both
expressions lead to the same high-energy limit. In this limit,
$\langle \psi_f^{(0)} | \psi_i \rangle$ as well as $\langle \psi_f^{(0)} |
V_{pe} \psi_i \rangle$ both become proportional to $\langle \phi_{\epsilon_f}
(\mathbf{r_1}) | \psi_i (\mathbf{r_1}, \mathbf{r_2} =0) \rangle$ (where $\phi$
denotes the one-electron state of the shaken electron and $\epsilon$ its
energy) with different prefactors.
Since they appear equally in the numerator and the denominator of (\ref{e14})
they cancel out, leading to the same high-energy limit. The same is true for
{\em all} energies if, as in SCR, product wavefunctions are used for the
initial state, or if the PEAK approximation is employed
\cite{pat01b,sch02,sch03}, i.e.\ if the absorption of the photon is assumed to
happen always at the nucleus at any excess energy $E$.
In this case one arrives at the ``natural''
definition (equation (8) of SCR)
\begin{equation}
\label{e15}
P_{f,abs}^{S} = \theta \left( E - \epsilon_f \right) \; | \langle
\phi_{\epsilon_f} | \phi_i \rangle |^2 
\end{equation}
(where $\theta$ is the unit step function),
i.e.\ the overlap of two one-electron wavefunctions. (In the case of using the
PEAK approximation for a correlated initial state $\phi_i \equiv \psi_i
(\mathbf{r_1}, \mathbf{r_2} =0) / \langle \psi_i(\mathbf{r_1}, \mathbf{r_2}
=0) | \psi_i(\mathbf{r_1}, \mathbf{r_2} =0) \rangle^{1/2}$.)

As argued above, the successful application of the SCR method showing excellent
agreement with experiment and {\em ab-initio} calculations suggests the
adoption of equation (\ref{e14}) as a good ``operational'' definition of
shake-off. In this spirit, shake-off may be phrased vaguely as the part of
the double ionization that is absent in a full collision (due to the
orthogonality of initial and final states), or, more precisely,
that part which does not involve an electron-electron interaction explicitly
(of course, interaction is implicit in the correlation present in the
initial state, without which there would be no shake-off). It should be
noted that the quality of this definition depends on the observation that there
is very little interference between shake-off and knock-out, as discussed
above. The fact that this is not strictly true leads to some problems e.g.\
at very low energies, where this separation would lead to a linear dependence
of the double ionization cross section on the excess energy, in contrast to
the well-known Wannier threshold power law \cite{wan53}.

On the other hand, one might want to adopt a maybe more ``physical'' definition
of shake-off on the basis of the intuitive picture of a time-dependence of
the effective one-electron potential the ``shaken'' electron feels. For
asymptotically high energies, it is the sudden change of this potential that
leads to a relaxation of the electron which is not in an eigenstate anymore
after the potential has changed. In this sense, the change in the effective
potential does not occur suddenly anymore at finite energies, but rather
over a timescale given by the velocity of the outgoing photoelectron.
This is the basis for the definition of shake-off adopted in \cite{tho84},
where an expression has been derived from time-dependent perturbation theory,
and also the rationale behind the somewhat {\em ad-hoc} formulation used in
\cite{pat01b} which is motivated by a Rosen-Zener-like expression for
diabatic transition probabilities familiar from ion-atom collisions. It should
be noted that, in contrast to (\ref{e14}), both of these expressions show an
exponential decrease towards threshold, so that the Wannier threshold law is
recovered. While in the SCR-expression the probability to be shaken into a
specific final state does not depend on the rate of change of the potential
(i.e.\ the velocity $v$ of the photoelectron) as long as it is energetically
allowed, this is different in \cite{tho84,pat01b} where these
probabilities depend exponentially on $v^2$ and $v$, respectively. In view
of these differences, it is surprising to see that the numerical values
resulting from these different definitions are in fact rather similar, as is
demonstrated in figure 1.
\begin{figure}[hbt]
\centerline{\psfig{figure=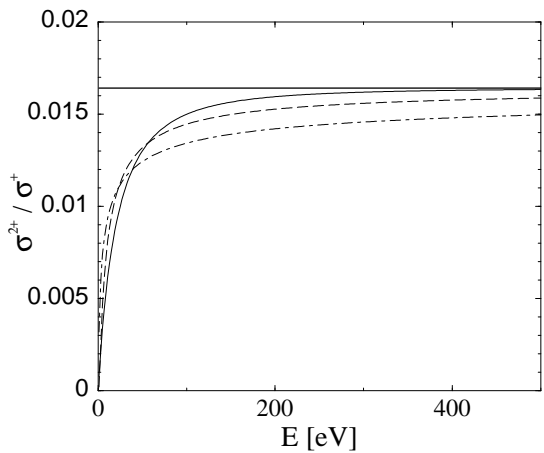,height=6cm}}
\caption[]{Comparison of different shake-off expressions for the case of the
helium ground state. Solid line: Schneider et al.\ \cite{sch02}, dashed:
Thomas \cite{tho84}, dot-dashed: Pattard and Burgd\"orfer \cite{pat01b};
the thick solid line shows the asymptotic $E \to \infty$ limit.}
\end{figure}
This leads us to suggest (\ref{e14}) as a good working definition
of shake-off at finite energies. It agrees qualitatively
with other natural and maybe more ``physical'' definitions and, moreover,
it has been shown to lead to a very good approximate separation of mechanisms
into shake-off and knock-out. It is easy to calculate and, as compared to
\cite{tho84,pat01b}, has the advantage that it does not contain any free
parameters (such as a characteristic range of interaction as in \cite{tho84}
or an effective impact parameter as in \cite{pat01b}). However, one should
always keep in mind that the significance of (\ref{e14}) as an independent
physically meaningful
quantity is limited, as discussed, e.g., in connection with the behaviour of
the cross section near threshold.

At this point, a short comment on a comparison with other available
definitions of shake-off seems to be in place. So far we have given a
comparison with \cite{tho84,pat01b} only, and used the qualitative agreement
with the values calculated from the expressions given there as an argument in
favour of the current definition. However, as argued before, some other
definitions found in the literature lead to shake-off values rather
different from the present. The shake-off calculated from many-body
perturbation theory (MBPT) \cite{hin93},
e.g., is found to have a completely different behaviour.
Its shape (as a function of energy) as well as the high-energy limit reached,
even the answer to the question whether it is the dominant process at high
energies or not,
depend on the choice of the gauge in which the corresponding diagram is
calculated. This is not too surprising in view of the fact that only the sum
of all first-order diagrams (shake-off plus ground-state correlation plus
two-step one) has a well-defined and gauge invariant meaning. Hence, the
meaning of shake-off is well defined within MBPT, however it is not claimed
to have any independent physical meaning of its own. In this sense, it is
not a helpful quantity if one wishes to discuss approximate physical
mechanisms. Another definition, originally formulated by {\AA}berg
\cite{abe69}, has recently been used by Shi and Lin \cite{lin02} to calculate
shake-off double ionization of the helium ground state. Their result for the
double-to-single cross section ratio is
found to be significantly larger than the latest experimental data for the
{\em total} ratio, i.e.\ including all possible decay routes (e.g.\ shake-off
and knock-out). From \cite{lin02} it is not entirely clear how much of this
``overshooting'' has to be attributed to a poor choice of the ground state
wavefunction (leading to an asymptotic high-energy limit which is somewhat
too large) and how much would still be observed using a more accurate
initial state. Assuming that this effect persists it would be obvious that
again no physical meaning can be ascribed to this definition of shake-off,
since shake-off alone would already be larger than the sum of all mechanisms.
In any
case, a further discussion would have to await a corresponding formulation
of ``non-shake-off'', since it is only the sum of all possible ionization
mechanisms which can directly be compared with experiment. Finally, Kheifets
\cite{khe01} has proposed a definition of shake-off where the diagonal part
of the $T$-matrix contribution to the convergent close-coupling model is
absorbed into shake-off. In his calculations for helium,
it was found that with this
definition of shake-off the total cross section ratio approaches the shake-off
value quickly, and non-shake-off becomes negligible at about 100 eV excess
energy. However, the last panel of figure 2 of \cite{khe01} shows that for
lower energies
shake-off alone again exceeds the total ratio. Hence, once more one has to
conclude that the meaning of shake-off as defined in \cite{khe01} as an
independent physical mechanism is limited.

In summary, we have argued that no unique definition for shake-off at finite
energies exists.
Nevertheless, we propose equation (\ref{e14}) as a good ``operational''
definition. Clearly, when shake-off at finite energies is
discussed
(in particular in the sense of an approximate physical mechanism in connection
with experiments), care has to be taken of the precise meaning of the term,
i.e.\ its actual definition adopted in
each case. In addition to our discussion of shake-off, we have indicated a
way towards a rigorous derivation of the SCR ansatz for calculation of double
ionization by relating it to a perturbation expansion starting from a full
quantum mechanical point of view.

T.P. would like to thank Joachim Burgd\"orfer for stimulating his interest
in the problem of single-photon multiple ionization, as well as many very
helpful discussions.

\end{document}